\documentclass[conference]{IEEEtran}

\IEEEoverridecommandlockouts

\usepackage{color}
\usepackage{amsfonts,amsmath,amssymb,amsthm}
\usepackage{latexsym,amscd}
\usepackage{amsbsy}
\usepackage{graphicx,multirow,bm}
\usepackage{algorithm}
\usepackage{algorithmicx}
\usepackage{algpseudocode}
\usepackage{xcolor}
\usepackage{tikz}
\usepackage{diagbox}
\usepackage{multicol}
\usepackage{arydshln}
\usepackage{booktabs}
\usepackage{bbm}
\usepackage{caption}
\usepackage{times}
\usepackage{amsmath}
\usepackage{cases}
\usepackage{subfigure}
\usepackage{cite}
\usepackage[top=0.7in,right=0.64in,left=0.625in,bottom=1.1in]{geometry} 

\newtheorem{theorem}{Theorem}
\newtheorem{lemma}{Lemma}

\newtheorem{definition}{Definition}

\newtheorem{remark}{Remark}

\begin{document}
	
	\title{Error Correcting Codes for Segmented  Burst-Deletion Channels}	
\author
{
	\IEEEauthorblockN{
		Yajuan Liu and Tolga M. Duman}\\
	\IEEEauthorblockA{EEE Department, Bilkent University, Ankara, Turkey\\}
	Email: yajuan.liu@bilkent.edu.tr, duman@ee.bilkent.edu.tr
	
	\thanks{This work was funded by the European Union through the ERC Advanced
		Grant 101054904: TRANCIDS. Views and opinions expressed are, however,
		those of the authors only and do not necessarily reflect those of the European
		Union or the European Research Council Executive Agency. Neither the
		European Union nor the granting authority can be held responsible for them.}
}

\maketitle
\begin{abstract}
	We study  \textit{segmented burst-deletion channels}  motivated by the observation that synchronization errors commonly occur in a bursty manner in real-world settings.
	In this channel model, transmitted sequences are implicitly divided into non-overlapping segments, each of which may experience at most one burst of deletions. 
	In this paper, we develop error correction codes for segmented burst-deletion channels over arbitrary alphabets under the assumption that each segment may contain only one burst of $t$-deletions. The main idea is to encode the input subsequence corresponding to each segment using existing one-burst deletion codes, with additional constraints that enable the decoder to identify segment boundaries during the decoding process from the received sequence. The resulting codes achieve redundancy that scales as $O(\log b)$, where $b$ is the length of each segment.
\end{abstract}

\section{Introduction}\label{sec:Introduction}
Synchronization error channels with insertions and deletions have garnered significant interest motivated by different applications, including magnetic data storage, DNA data storage, and document synchronization. 
In practical systems, the inherent timing mismatch between the device reading the data and the data layout typically leads to errors that occur intermittently, with noticeable gaps between error events. In other words, once an error event occurs, be it an insertion, deletion, or even a burst of insertions/deletions, it is less likely that another error will follow immediately.
Building on this observation,  Liu and Mitzenmacher proposed the  \textit{segmented edit channel}, for which transmitted sequences are implicitly divided into non-overlapping segments, with at most one insertion or deletion occurring in each segment \cite{liu2010codes}.

In a segmented edit channel, the input sequence of length $n=b\gamma$ is divided into $\gamma$ disjoint segments, each containing $b$ consecutive symbols. We assume that there is at most one insertion or deletion event per segment. If the segment boundaries were known, simply applying a
Varshamov-Tenengolts (VT) code \cite{varshamov1965code} to  each segment would suffice. However, the boundaries are not known, requiring different designs. As the first approach on segmented edit channels, Liu and Mitzenmacher \cite{liu2010codes} provided several conditions for these channels, and designed segmented insertion and deletion error correcting codes (ECCs) with $b\le9$. These codes, however, cannot be easily extended to larger values of $b$ due to increased computational complexity.
Subsequently, Wang \textit{et al.} \cite{wang2013capacity} developed a theoretical characterization and a practical coding approach  for segmented deletion channels by concatenating outer  low density parity check (LDPC) codes  with inner marker codes.
In \cite{abroshan2018coding}, the segmented edit channels were extended to non-binary alphabets, and code constructions for segmented insertion, deletion, and insertion/deletion (indel) channels were designed. Specifically, the authors propose binary segmented ECCs, which are capable of handling $1$-insertion ($1$-ins), $1$-deletion ($1$-del) or $1$-indel with redundancies of $\log(b+1)+2$, $\log(b+1)+2.5$ and $\log(b+1)+7$ bits per segment,  while the redundancies of $q$-ary ECCs are  $\log b+6-2\log 3$, $\log b+8$ and $\log b+16$ bits, respectively.\footnote{Unless stated otherwise, all logarithmic operations are base $2$.}
These constructions are based on subsets of VT codes chosen with predetermined prefixes and/or suffixes.
Building upon this work, Jiao \textit{et al.} improved the redundancy of ECCs over binary segmented deletion channels to $\log(b+1)+2-\log 1.5$ bits per segment  and developed systematic encoding algorithms in \cite{jiao2022on}.
After that, focusing on quaternary sequences, Cai \textit{et al.} \cite{cai2021coding} designed segmented edit ECCs with local GC-balance constraints to reduce the probability of error in DNA storage systems. Specifically, they provide quaternary segmented ECCs for correcting $1$-indel  with a redundancy of $\log b+6\log 3+6$ bits per segment. Yan \textit{et al.} \cite{yan2023asegment} introduced segmented ECCs for
correcting $1$-insertion/deletion/substitution (edit) with $2\log b'+14$ bits of redundancy per segment by incorporating markers into VT codewords, where $b=b'+\lceil \log b'\rceil +7$.
More recently, the work in \cite{li2024marker} constructed binary segmented ECCs for correcting $1$-indel and $1$-edit with redundancies of $\log(b-6)+7$ and $\log(b-9)+10$ bits per segment.

\begin{table*}[tb]
	\caption{The comparisons of some known segmented ECCs}
	\label{tab:comparison}
	\centering
		\begin{tabular}{ccccccc}
			\hline
			Alphabet & Error type & Redundancy for each segment & Ref.\\
			\hline
			\multirow{10}{*}{$q=2$} & $1$-del. & \multirow{2}{*}{/} & \multirow{2}{*}{\cite{liu2010codes}}\\
			&$1$-ins. &       \\
			\cline{2-5}
			& $1$-del. &  $\log(b+1)+2$   &  \multirow{3}{*}{\cite{abroshan2018coding}}\\
			&$1$-ins. &  $\log(b+1)+2.5$  &               \\
			&$1$-indel &  $\log(b+1)+7$   &\\
			\cline{2-5}
			&$1$-del. &           $\log(b+1)+2-\log 1.5$ &   \multirow{2}{*}{\cite{jiao2022on}}      \\
			&$ 1$-ins. &           $\log(b+1)+2.5$     &      \\
			\cline{2-5}
			&$ 1$-indel. &  $\log(b-6)+7$   & \multirow{2}{*}{\cite{li2024marker}}\\
			&$1$-edit &  $\log(b-9)+10$   & \\
			\cline{2-5}
			\hline
			\multirow{2}{*}{$q=4$}&$ 1$-indel. &    $\log b+6\log 3+6$ & \cite{cai2021coding}                     \\
			\cline{2-5}
			&$ 1$-edit &           $2\log b'+14,~b=b'+\lceil \log b'\rceil +7$   &\cite{yan2023asegment}          \\
			\hline
			\multirow{5}{*}{$q>2$}& $ 1$-del. & $\log b+6-2\log 3$ &  \multirow{3}{*}{\cite{abroshan2018coding}} \\
			&$ 1$-ins. &            $\log b+8$              \\
			&$1$-indel &    $\log b+16$                      \\
			\cline{2-5}
			& $1$-burst of $\le t_1$-del/$\le t_2$-ins. &$b\log b/u+\log b$, $n/b=t_1+t_2+2$&\multirow{2}{*}{\cite{yi2024error}}\\
			& $1$-burst of $\le t_1$-del/$\le t_2$-ins. &$((\lambda+2u)b\log b)/(2u(\lambda+1))+\log b$ &  \\
			\hline
			$q=2$&\multirow{2}{*}{$ 1$-burst of $t$-del.}& $\log b+o(\log b)+3$& Thm. \ref{thm:binary}\\
			$q>2$ & &$\log b+o(\log b)+5\log q-4\log (q-1)$ & Thm. \ref{thm:q-ary}\\
			\hline
	\end{tabular}
\end{table*}


In some practical applications such as racetrack memory and high density magnetic recording channels, data may be corrupted by bursts of insertions/deletions  \cite{parkin2008magnetic,mazumdar2011channels},
which means that the insertion/deletion errors may occur at consecutive positions and the maximal length of consecutive errors is limited \cite{zhang2015hifi}.
Based on this observation, Yi \textit{et al.} investigate the \textit{segmented burst-indel channel}  for non-binary alphabets in \cite{yi2024error}, where at most one burst of indel  occurs per segment.
They construct  two ECCs by means of maximum distance separable (MDS) codes and  binary marker patterns for non-binary sequences with burst lengths proportional to the code length, which are capable of correcting one burst of at most $t_1$-del/$t_2$-ins in each segment. However, the redundancies of the resulting codes are $b\log b/u+\log b$ or  $((\lambda+2u)b\log b)/(2u(\lambda+1))+\log b$ bits, scaling on the order of $O(b\log b)$, where $1/u=(t_1+t_2+2)/b$ is roughly the ratio of the maximum length of burst insertions or deletions to the segment length, and $\lambda$ is a positive integer.

In this paper, we consider a burst-deletion channel model, in which the burst length is independent of the code length, rather than being proportional to it. That is, our schemes are capable of correcting one burst of $t$-del for any constant $t$ with respect to $b$, in contrast to prior work \cite{yi2024error}, which assumes $t=b/u-2$. Specifically, we construct segmented burst-deletion ECCs for $q$-ary ($q\ge 2$) sequences,  achieving a redundancy of $O(\log b)$. 
Our work is based on specific subsets of codes designed to handle a single burst of deletions.
By introducing some additional restrictions on the codewords of one burst-deletion ECCs,  the designed codes can identify the segment boundaries and correct one burst of $t$-del per segment.

	The remainder of this paper is organized as follows.
Section \ref{sec:preliminary} reviews some necessary preliminaries. In Section \ref{sec:deletion}, we  develop the segmented burst-deletion ECCs for both binary and non-binary sequences. Finally,
concluding remarks are presented in Section \ref{sec:conclusions}.

	\section{Preliminaries}\label{sec:preliminary}
In this section, we review the $q$-ary ($q\ge 2$) ECCs in \cite{song2023nonbinary} and \cite{song2024new}, which can correct one burst of $t$-del with $\log n +8\log\log n+o(\log\log n)$ bits of redundancy, where $n$ is the codelength.  We will employ these codes to construct segmented burst-deletion ECCs for binary and non-binary sequences in Section \ref{sec:deletion}.

We denote  vectors and sets by bold lowercase and  calligraphic letters, respectively, \textit{e.g.},  $\mathbf{a}$ and $\mathcal{A}$, and a sequence of length $t$ formed by $a\in\{0,1\}$ as $a^t$.
The length of $\mathbf{a}$ and size of $\cal{A}$ are denoted by $|\mathbf{a}|$ and $|\mathcal{A}|$, respectively.
For two non-negative integers $a$ and $b$ with $a < b$, define two ordered sets $\{a,a+1,\dots,b-1\}$ and $\{a,a+1,\dots,b\}$ by $[a,b)$ and $[a,b]$, respectively.
We define $\Sigma_q=\{0,1,\dots,q-1\},q\ge 2$, and for a  sequence $\mathbf{x}=(x_1,x_2,\dots,x_n)\in\Sigma_q^n$ and any two positive integers $i<j$, we write $\mathbf{x}(i:j)=(x_i,x_{i+1},\dots, x_j)$.
Furthermore, for a set $\mathcal{L}=[i,j]\subseteq[1,n]$, we define $\mathbf{x}_{\mathcal{L}}\triangleq\mathbf{x}(i:j)$.
Finally, we show by $\mathcal{B}_t^D$ the set of sequences obtained from $\mathbf{x}$ by deleting  $t$ consecutive symbols, i.e., one burst of $t$-del, where $D$ represents \textit{deletion} for brevity.

\subsection{Channel Model}
For any sequence $\mathbf{x}=(x_1,x_2,\dots,x_n)\in\Sigma_q^n$, if \textit{one burst of $t$-del} occurs in the $i$-th coordinate of $\mathbf{x}$, 	where $i\in[1,n-t]$, we obtain a length $n-t$ sequence $\mathbf{x}'\in \mathcal{B}_{t}^D(\mathbf{x})$ as
\begin{equation*}
	\mathbf{x}'=(x_1,x_2,\dots,x_{i-1},x_{i+t},\dots,x_n)\in\Sigma_q^{n-t}.
\end{equation*}

In the segmented burst-deletion channel, the input sequence is denoted by $\mathbf{x}=(x_1,x_2,\dots,x_n)\in\Sigma_q^n$, which is divided into $\gamma$ segments, each with length $b$. That is, $n=b\gamma$. The $i$-th segment of $\mathbf{x}$ is denoted by $\mathbf{x}_i=\mathbf{x}_i(1:b)=(x_{i,1},x_{i,2},\dots,$ $x_{i,b})=\mathbf{x}((i-1)b+1:ib)$ for $i\in[1,\gamma]$.
The channel output sequence, denoted by $\mathbf{y}=(y_1,y_2,\dots,y_m)$ with $m\le n$, is obtained from $\mathbf{x}$ by deleting at most one burst of $t$ symbols in each segment. This means that there are at most $\gamma$ bursts of $t$-del in $\mathbf{y}$, and roughly speaking, they are separated from each other due to the segmented nature of deletions.
Define 
\begin{eqnarray*}
	\mathcal{B}_{\gamma,t}^D(\mathbf{x})\triangleq\{\mathbf{y}\in\Sigma_q^m:\mathbf{y} \text{~is obtained from~} \mathbf{x} \text{~by deleting at }\\
	\text{most one burst of~} t \text{~symbols per segment}\}.
\end{eqnarray*} 
We assume that the decoder knows $\gamma$ and $b$, but not the segment boundaries after the deletion events.

\begin{definition}
	Let $n=b\gamma$ for two positive integers $b$ and $\gamma$, and $\mathcal{C}$ be a subset of $\Sigma_q^n$ with $|\mathcal{C}|\ge 2$, where $q\ge 2$.  We call $\mathcal{C}$ a \textit{segmented burst of $t$-del ECC} if for any two distinct sequences $\mathbf{x},\mathbf{x}'\in\mathcal{C}$,
	$
	\mathcal{B}_{\gamma,t}^D(\mathbf{x})\cap \mathcal{B}_{\gamma,t}^D(\mathbf{x}')=\varnothing.
	$
\end{definition}

For a segmented burst of $t$-del ECC $\mathcal{C}\subseteq \Sigma_q^n$, we define the \textit{rate}
of	$\mathcal{C}$ as $R=(\log {|\mathcal{C}|})/n$.
We consider a segment by segment encoding.
If the number of codewords for each segment is $M$, then the rate of $\mathcal{C}$ is
\begin{eqnarray}\label{eqn:rate}
	R={\log {|\mathcal{C}}|\over n}={\log M^{\gamma}\over n}={1\over b }\log  M.
\end{eqnarray}

\subsection{One burst of $t$ deletion ECCs}\label{sec:burst}
One burst of $t$-del ECCs are developed in \cite{lenz2020optimal,song2023nonbinary,song2024new,schoeny2017codes,wang2024nonbinary}.
In this subsection, we review the codes proposed in \cite{song2023nonbinary} and \cite{song2024new}, which developed one burst of $t$-del ECCs for binary and non-binary $(\mathbf{p},\delta)$-dense sequences, respectively.

\begin{definition}\label{def:seq}
	Let $\delta\le n$ be an integer and $\mathbf{p}$ be a pattern.
	A $q$-ary sequence $\mathbf{x}\in\Sigma_q^n$ is said to be a \textit{$(\mathbf{p},\delta)$-dense sequence}
	if there is at least one $\mathbf{p}$ in any length $\delta$ substring of $\mathbf{x}$.
\end{definition}

	\begin{remark}\label{rem:burst}
			The works in \cite{lenz2020optimal} and \cite{song2024new} construct efficient encoding and decoding algorithms with only $3$ bits of redundancy in time $O(n)$ for binary $(\mathbf{p}=0^t1^t,\delta=t2^{2t+1}\lceil\log n\rceil)$-dense sequences, and one bit of redundancy in time $O(n\log n)$ for $q$-ary $(\mathbf{p}=0^t1^t,\delta=2tq^{2t}\lceil\log n\rceil)$-dense sequences ($q>2$), respectively.
	\end{remark}

	
	\begin{lemma}[\cite{song2024new,sima2020syndrome}]\label{lemma:burst-t}
	For any $\mathbf{x}\in\Sigma_q^n$,	there exists a function $h: \Sigma_q^n \rightarrow \Sigma_q^{4\log_q n+o(\log_q n)}$, such that given $h(\mathbf{x})$ and any $\mathbf{x}'\in \mathcal{B}_t^D(\mathbf{x})$,
	one can uniquely recover $\mathbf{x}$ in time $O(n^3q^t)$.
\end{lemma}

Set $\mathbf{p}=0^t1^t$ and $\delta=2tq^{2t}\lceil\log n\rceil$. 	For a $(\mathbf{p},\delta)$-dense sequence $\mathbf{x}\in\Sigma_q^n$, assume that there are $m$ $\mathbf{p}$'s in $\mathbf{x}$, then $\mathbf{x}$ can be rewritten  as
$\mathbf{x}=(\mathbf{x}'_00^t1^t\mathbf{x}'_10^t1^t\cdots \mathbf{x}'_{m-1}0^t1^t\mathbf{x}'_m)$,
where $\mathbf{x}'_i,i\in[1,m]$ is a substring of $\mathbf{x}$, which does not contain $\mathbf{p}$. Moreover, $|\mathbf{x}'_0|,|\mathbf{x}'_m|\le \delta-2t$ and $|\mathbf{x}'_i|\le \delta -4t-1,~i\in[1,m-1]$ according to Definition \ref{def:seq}.

Define the \textit{$\mathbf{p}$-indicator vector} of $\mathbf{x}$ by 
$$\mathbbm{1}_{\mathbf{p}}(\mathbf{x})=(\mathbbm{1}_{\mathbf{p}}(\mathbf{x})_1,\mathbbm{1}_{\mathbf{p}}(\mathbf{x})_2,\dots,\mathbbm{1}_{\mathbf{p}}(\mathbf{x})_n)\in\Sigma_2^n,$$
where $\mathbbm{1}_{\mathbf{p}}(\mathbf{x})_i=1$ if $(x_i,x_{i+1},\dots,x_{i+2t-1})=\mathbf{p}$, otherwise it is $0$.
Let $a_0(\mathbf{x})=\sum_{i=1}^n \mathbbm{1}_{\mathbf{p}}(\mathbf{x})_i \pmod 4$ and $a_1(\mathbf{x})=\sum_{i=1}^n i\cdot\mathbbm{1}_{\mathbf{p}}(\mathbf{x})_i \pmod {2n}$.

Let $\rho=\delta+t$ for binary sequences, and $\rho=3\delta=6tq^{2t}\lceil\log n\rceil$ for the $q$-ary ($q>2$) case. Define
\begin{eqnarray}\label{eqn:interval}
	\mathcal{L}_j\triangleq 	\!\left\{\begin{array}{ll}
		\!\!\!	[(j-1)\rho+1,(j+1)\rho], \text{for~} j\in[1,\lceil n/\rho\rceil -2],\\
		\!\!\!	{[(j-1)\rho+1,n]},\text{~~~~~~~~for~} j=\lceil n/\rho\rceil -1.
	\end{array}\right.
\end{eqnarray}

\begin{lemma}[\cite{song2023nonbinary,song2024new}]\label{lemma:burst}
	For any $(\mathbf{p},\delta)$-dense sequence $\mathbf{x}\in\Sigma_q^n$, $q\ge 2$, let $h(\mathbf{x})$ be the function constructed in Lemma \ref{lemma:burst-t}, and $\mathcal{L}_j,j\in[1,\lceil n/\rho\rceil),$ be the sets defined in \eqref{eqn:interval}.
	Let
	\begin{eqnarray*}\label{eqn:fx}
		f(\mathbf{x})=\left(a_0(\mathbf{x}),a_1(\mathbf{x}),\bar{h}^{(0)}(\mathbf{x}),\bar{h}^{(1)}(\mathbf{x})\right),
	\end{eqnarray*}
	where
	\begin{eqnarray*}\label{eqn:bar-h}
		\bar{h}^{(\ell)}(\mathbf{x})=\sum_{j\in[1,\lceil n/\rho\rceil)\atop j\equiv \ell \bmod 2} h(\mathbf{x}_{\mathcal{L}_j}) \pmod{q^{N}}, \quad\ell\in\{0,1\},
	\end{eqnarray*}
	where $N=4(\log_q 2\rho)+o(\log_q 2\rho)$.
	
	Then, given $f(\mathbf{x})$ and any $\mathbf{x}'\in \mathcal{B}_t^D(\mathbf{x})$,
	one can uniquely recover $\mathbf{x}$ with $\log n+8\log\log n+o(\log\log n)+r_{q,t}$ bits of redundancy, computable in time $O(n(\log n)^3q^{7t})$, where $r_{q,t}$ is a constant with respect to $n$.
\end{lemma}

In what follows, we refer to one burst of $t$-del ECCs summarized in Lemma \ref{lemma:burst} as \textit{SK codes} when $q=2$ \cite{song2023nonbinary}, and as \textit{SKQ codes} when $q>2$ \cite{song2024new}.

\section{ECCs for Segmented Burst-Deletion Channels}\label{sec:deletion}
In this section, we develop ECCs for segmented burst-deletion channels. 
Our approach is inspired by the work of \cite{abroshan2018coding}, where the authors design ECCs for segmented edit channels by first encoding each segment via a VT code, and then  selecting specific subsets by fixing the prefixes/suffixes of codewords. 
In a similar fashion, we encode each segment into a codeword of SK codes or SKQ codes, and then
fix certain components (such as, the first or the $t+1$-th bit, rather than prefixes/suffixes) of each codeword (see Subsections \ref{sub-a} and \ref{sub-b} for details). Following this,
the segment boundaries become identifiable during the decoding process.
We assume that for each segment with length $b$, the number of deleted symbols $t$ satisfies $2t<b$, which is a realistic assumption for practical communication channels.

Let $\mathcal{P}_q$ be the set of $q$-ary $(\mathbf{p},\delta)$-dense sequences, where $\mathbf{p}=0^t1^t$, $\delta=2tq^{2t}\lceil\log n\rceil$. Recall that the function $f(\mathbf{x})$ with syndrome $a_0(\mathbf{x}),a_1(\mathbf{x}),\bar{h}^{(0)}(\mathbf{x}),\bar{h}^{(1)}(\mathbf{x})$ constructed in Lemma \ref{lemma:burst} can correct one burst of $t$-del for $\mathbf{x}\in\mathcal{P}_q$.
For convenience, we denote 
$\alpha=a_0(\mathbf{x}),\beta=a_1(\mathbf{x}),\eta=\bar{h}^{(0)}(\mathbf{x})$ and $\zeta=\bar{h}^{(1)}(\mathbf{x})$ in the following.

\subsection{Binary segmented burst-deletion ECCs}	\label{sub-a}
Let $\mathbf{x}$ be a binary $(\mathbf{p},\delta)$-dense sequence written as $\mathbf{x}=\mathbf{x}_1\mathbf{x}_2\cdots\mathbf{x}_{\gamma}\in\mathcal{P}_2$, where $\mathbf{x}_i$ is the $i$-th segment of $\mathbf{x}$ with $|\mathbf{x}_i|=b,i\in[1,\gamma]$, and $b>2t$. For four non-negative integers $\alpha\in[0,3],\beta\in[0,2b),\eta\in[0,2^{N}),\zeta\in[0,2^{N})$, where $N=4(\log 2\rho)+o(\log 2\rho)$ and $\rho=\delta+t$, define
\begin{eqnarray}
	\mathcal{S}^{(0)}_{\alpha,\beta,\eta,\zeta}&\triangleq&\left\{\mathbf{s}\in\mathcal{P}_2\subseteq\Sigma_2^b: \mathbf{s} \text{ satisfies $f(\mathbf{s})$ with }\right.\nonumber\\
	&&\left.\qquad s_{1}=0,s_{t+1}\ne s_{2t},s_{b-t+1}=s_{b} \right\},\label{eqn:S0}\\
	\mathcal{S}^{(1)}_{\alpha,\beta,\eta,\zeta}&\triangleq&\left\{\mathbf{s}\in\mathcal{P}_2\subseteq\Sigma_2^b:\mathbf{s} \text{ satisfies $f(\mathbf{s})$ with }\right.\nonumber\\
	&&\left.\qquad s_{1}=1,s_{t+1}\ne s_{2t},s_{b-t+1}=s_{b}\right\}.\label{eqn:S1}
\end{eqnarray}

We observe that the sets $\mathcal{S}^{(0)}_{\alpha,\beta,\eta,\zeta}$ and $\mathcal{S}^{(1)}_{\alpha,\beta,\eta,\zeta}$ are two subsets of SK codes in Lemma \ref{lemma:burst}. Therefore, one burst of $t$-del for any segment $\mathbf{x}_i,i\in[1,\gamma],$ can be corrected if we can identify the starting positions of each segment.
To achieve this, we restrict the first component of SK codes to two different symbols and introduce additional conditions to the $t+1,2t,b-t+1,b$-th components to facilitate the identification of segment boundary when a burst-deletion occurs.

We select the parameters $\alpha\in[0,3],\beta\in[0,2b),\eta,\zeta\in[0,2^N)$ that maximize the number of codewords,
that is,
\begin{eqnarray*}
	(\alpha_0,\beta_0,\eta_0,\zeta_0)=\mathop{\arg\max}\limits_{\substack{0\le \alpha< 4,0\le \beta < 2b,\\ 0\le \eta,\zeta < 2^{N}}}|\mathcal{S}^{(0)}_{\alpha,\beta,\eta,\zeta}|,
\end{eqnarray*}
and
\begin{eqnarray*}
	(\alpha_1,\beta_1,\eta_1,\zeta_1)=\mathop{\arg\max}\limits_{\substack{0\le \alpha < 4,0\le \beta < 2b,\\ 0\le \eta,\zeta < 2^{N}}}|\mathcal{S}^{(1)}_{\alpha,\beta,\eta,\zeta}|.
\end{eqnarray*}
Then, 
\begin{eqnarray*}
	M=\min(|\mathcal{S}^{(0)}_{\alpha_0,\beta_0,\eta_0,\zeta_0}|,|\mathcal{S}^{(1)}_{\alpha_1,\beta_1,\eta_1,\zeta_1}|).
\end{eqnarray*}

In the following, we choose $M$ sequences from each of $\mathcal{S}^{(0)}_{\alpha_0,\beta_0,\eta_0,\zeta_0}$ and $\mathcal{S}^{(1)}_{\alpha_1,\beta_1,\eta_1,\zeta_1}$  to encode each segment such that the cardinality of the resulting codes is the largest. Denote the two sets of $M$ sequences as $\mathcal{S}^{(0)}$ and $\mathcal{S}^{(1)}$, respectively.

\textbf{Encoding:} The encoding is performed segment by segment starting from the first one.
The sequence $\mathbf{x}\in\mathcal{P}_2$ is encoded as concatenation of the selected codewords in $\mathcal{S}^{(0)}$ and $\mathcal{S}^{(1)}$. Specifically, 	we encode the first segment $\mathbf{x}_1$ to a sequence of $\mathcal{S}^{(0)}$. Then,
the $i$-th segment $\mathbf{x}_i,i\in[2,\gamma],$  of length $b$ is encoded by choosing a sequence from $\mathcal{S}^{(1)}$ if the last bit of  $\mathbf{x}_{i-1}$ is $0$, otherwise, a sequence from $\mathcal{S}^{(0)}$ is chosen.

\textbf{Decoding:} The received sequence is decoded segment by segment. 
In the sequel, assume that $\mathbf{x}_{i-1}$ has been recovered successfully, then the syndrome of $\mathbf{x}_i$ is known, i.e., $f(\mathbf{x}_i)$ with either $\alpha_0,\beta_0,\eta_0,\zeta_0$ or $\alpha_1,\beta_1,\eta_1,\zeta_1$. Denote   the  starting position of $\mathbf{x}_i$ by $p_i+1$.

Denote the received sequence by $\mathbf{y}$ and the $i$-th segment $\mathbf{y}_i=\mathbf{y}(p_i+1:p_i+b)$ as an estimate of $\mathbf{x}_i,i\in[1,\gamma]$. If a burst of $t$-del occurs at the $a$-th component of $\mathbf{x}_i$, the first $t$ symbols of $\mathbf{y}_i$ become $\hat{\mathbf{y}}_i(1:t)=\mathbf{x}_i(1:2t)\backslash\mathbf{x}_i(a:a+t-1)$ for $1\le a\le t$ or $\hat{\mathbf{y}}_i(1:t)=\mathbf{x}_i(1:t)$ for $a>t$.
Particularly, when $a=1$, $\hat{\mathbf{y}}_i(1:t)=\mathbf{x}_i(t+1:2t)$.	
Note that it is impossible for $\hat{\mathbf{y}}_i(1:t)$ to be regarded as the last $t$ bits of $\mathbf{x}_{i-1}$ since $\hat{\mathbf{y}}_i(1:t)\ne \mathbf{x}_{i-1}(b-t+1:b)$ following 
$x_{i,1}\ne x_{i-1,b}=x_{i-1,b-t+1}$ and $x_{i,t+1}\ne x_{i,2t},x_{i-1,b-t+1}=x_{i-1,b}$. 
Therefore, the starting position $p_i+1$ is the first or the $t$-th element of $\mathbf{x}_i$.
Consequently,
if  $f(\mathbf{y}_i)\ne f(\mathbf{x}_i)$, according to the syndrome function $f(\mathbf{x}_i)$ and $\mathbf{y}(p_i+1:p_i+b-t)$,  the $i$-th segment can be recovered
and the starting position of  $\mathbf{x}_{i+1}$ is found as $p_i+b-t+1$. 
Otherwise, we deduce that $\mathbf{x}_i=\mathbf{y}_i$ and determine the starting position of $\mathbf{x}_{i+1}$, as described in Lemma \ref{fx}.

\begin{lemma}\label{fx}
	Let a binary $(\mathbf{p},\delta)$-dense sequence $\mathbf{x}=\mathbf{x}_1\mathbf{x}_2\cdots\mathbf{x}_{\gamma}$ be the input and $\mathbf{y}$ be the output of a segmented burst-deletion channel.
	Assume that the $i-1$-th segment of $\mathbf{x}$ is recovered successfully and the starting position  of $\mathbf{x}_i$ is $p_i+1$.	
	Denote $\mathbf{y}_i=\mathbf{y}(p_i+1:p_i+b)$. If $f(\mathbf{y}_i)=f(\mathbf{x}_i)$, then $\mathbf{x}_i=\mathbf{y}_i$ and the starting position of the $i+1$-th segment is $p_i+b+1$.
\end{lemma}
\begin{proof}
	Without loss of generality, suppose that the last bit of $\mathbf{x}_{i-1}$ is $1$, then the syndrome of $\mathbf{x}_i$ is $f(\mathbf{x}_i)$ with $(\alpha_0,\beta_0,\eta_0,\zeta_0)$. The decoding process for the case with the last bit being $0$, syndrome $f(\mathbf{x}_i)$ with  $(\alpha_1,\beta_1,\eta_1,\zeta_1)$ is similar.
	
	Following the encoding approach and \eqref{eqn:S0}, if there is no burst of $t$-del in $\mathbf{x}_i$, $x_{p_i+1}$ should be $0$.
	Therefore, if  $y_{p_i+1}=1$, it means that the initial $t$ bits of $\mathbf{x}_i$  are deleted.
	If $f(\mathbf{y}_i)=f(\mathbf{x}_i)$,  there exist two different sequences $\mathbf{x}_i$ and $\mathbf{y}_i=\mathbf{y}(p_i+1:p_i+b)=\mathbf{x}_i(t+1:b)\cup\hat{\mathbf{y}}_{i+1}(1:t)$  with
	the same syndrome $(\alpha_0,\beta_0,\eta_0,\zeta_0)$ and a common subsequence of length $b-t$  obtained by deleting $t$ consecutive symbols. This implies
	$\mathcal{B}_t^D(\mathbf{y}_i)\cap\mathcal{B}_t^D(\mathbf{x}_i)\ne \varnothing$. Hence, there is a contradiction.

		
In the following, we consider the case with $y_{p_i+1}=0$ by analyzing the following five cases shown in Figure \ref{fig}.
		\begin{figure}[htbp]
			\begin{center}
				\includegraphics[scale=0.45]{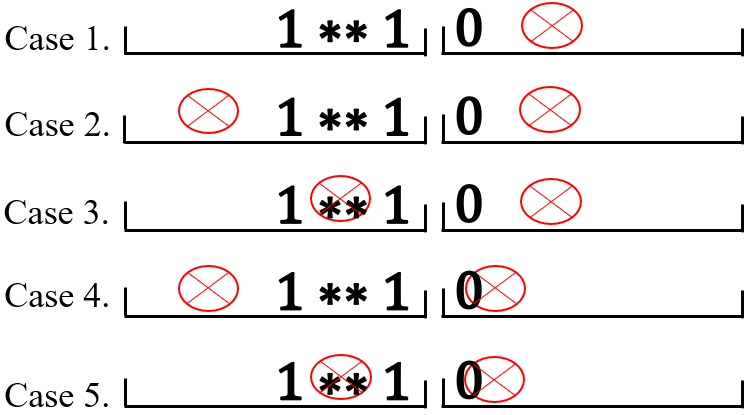}
			\end{center}
			\caption{The burst-deletion patterns for the decoding, where\\ $\otimes$ represents the position of one burst of $t$-del and empty space means the remainder data of the $i-1$-th and $i$-th segments.}
			\label{fig}
		\end{figure}
		
		\begin{itemize}
		\item [1.]	There is no burst of $t$-del in the $i-1$-th segment, and one burst of $t$-del in the $i$-th segment, which does not destroy the first bit of $\mathbf{x}_i$ (Case 1). Since we have recovered $\mathbf{x}_{i-1}$ and $\hat{\mathbf{y}}_i(1:t)$ can not be regarded as the last $t$ bits of $\mathbf{x}_{i-1}$, thus $y_{p_i+1}=x_{i,1}$. 
		Following this, if $f(\mathbf{y}_i)=f(\mathbf{x}_i)$,  there exist two different sequences $\mathbf{x}_i$ and $\mathbf{y}_i=\mathbf{y}(p_i+1:p_i+b)=\mathbf{x}_i(1:a-1)\cup\mathbf{x}_i(a+t:b)\cup\hat{\mathbf{y}}_{i+1}(1:t)$  with
		the same syndrome $(\alpha_0,\beta_0,\eta_0,\zeta_0)$ and a common subsequence of length $b-t$  obtained by deleting $t$ consecutive symbols, where $a\in[2,b-t+1]$ is the  position of burst-deletion in $\mathbf{x}_i$. Note that the fact of $\hat{\mathbf{y}}_{i+1}(1:t)\ne\mathbf{x}_i(b-t+1:b)$ ensures that  $\mathbf{x}_i$ and $\mathbf{y}_i$ are different.
		This implies that
		$\mathcal{B}_t^D(\mathbf{y}_i)\cap\mathcal{B}_t^D(\mathbf{x}_i)\ne \varnothing$, contradicting the fact that $f(\mathbf{x})$ with $(\alpha_0,\beta_0,\eta_0,\zeta_0)$ can correct one burst of $t$-del.

		\item [2.]
		There is one burst of $t$-del in the $i-1$-th segment and one in the $i$-th segment, respectively, which does not destroy the first bit of the latter (Cases 2-3).
		The burst-deletion of the $i-1$-th segment can be corrected successfully. After that, the proof is reduced to that of Case 1.
		
		\item [3.]
		There is one burst of $t$-del in the $i-1$-th segment and one affecting the initial $t$ bits of the $t$-th segment  (Cases 4-5).
		The burst-deletion of the $i-1$-th segment can be corrected successfully.
		Hence, $\mathbf{y}_i=\mathbf{y}(p_i+1:p_i+b)=\mathbf{x}_i(t+1:b)\cup\hat{\mathbf{y}}_{i+1}(1:t)$.
		If $f(\mathbf{y}_i)=f(\mathbf{x}_i)$,  there exist two different sequences $\mathbf{x}_i$ and $\mathbf{y}_i$  with
		the same syndrome $(\alpha_0,\beta_0,\eta_0,\zeta_0)$ and a common subsequence of length $b-t$  obtained by deleting $t$ consecutive symbols. This implies
		$\mathcal{B}_t^D(\mathbf{y}_i)\cap\mathcal{B}_t^D(\mathbf{x}_i)\ne \varnothing$. Hence, there is a contradiction.
	\end{itemize}
	
	Following the processes above, if $f(\mathbf{y}_i)=f(\mathbf{x}_i)$, then we know $\mathbf{y}_i=\mathbf{x}_i$, resulting in the starting position of $\mathbf{x}_{i+1}$ being $p_i+b+1$, completing the proof.	
\end{proof}

To summarize, according to \eqref{eqn:S0}, \eqref{eqn:S1} and the above encoding/decoding procedures, we obtain an ECC, which can identify the segment boundaries at the channel output and correct  one burst of $t$-del per segment.
In the sequel, we analyze the redundancy of the resulting codes.
From \eqref{eqn:S0},
there are $2^{b-1}$ binary sequences of length $b$ with the first bit $0$, out of which $12\times2^{b-5}$ are removed because they have the same symbols at the $t+1$-th and $2t$-th components and/or different symbols at the $b-t+1$-th and $b$-th components.
Each of these sequences belongs to exactly one of the sets $\mathcal{S}_{0,0,0,0}^{(0)},\dots,\mathcal{S}_{3,2b-1,2^N-1,2^N-1}^{(0)}$. Therefore, the size of the largest one among these $8b2^{2N}$ sets is lower bounded as
\begin{eqnarray*}
|\mathcal{S}^{(0)}|&\ge& {2^{b-1}-12\times2^{b-5}\over 4\cdot 2b\cdot 2^{2N}}\\
&=&{ 2^{b-3} \over 4\cdot 2b\cdot 2^{2N}}.
\end{eqnarray*}
A similar argument gives the same bound for $\mathcal{S}^{(1)}$ in \eqref{eqn:S1}, hence
\begin{eqnarray}\label{eqn:M}
M=\min(|\mathcal{S}^{(0)}|,|\mathcal{S}^{(1)}|)
\ge { 2^{b-3} \over 4\cdot 2b\cdot 2^{2N}}.
\end{eqnarray}

Taking logarithms of both sides, and applying \eqref{eqn:rate}, the rate of the designed codes is lower bounded by
\begin{eqnarray*}
R\ge 1-{|f(\mathbf{x})|\over b}-{3\over b},
\end{eqnarray*}
where $f(\mathbf{x})$ is defined in Lemma \ref{lemma:burst} for binary $(\mathbf{p},\delta)$-sequences. 

We see that the rate penalty with respect to  SK codes is at most $3/b$ bits per symbol.
Thus, the redundancy of the new codes is $\log b+8\log\log b+o(\log\log b)+r_{q,t}+3$ bits.
In other words, to correct one burst of $t$-del per segment over the segmented burst-deletion channel, we only need to add $3$ bits more redundancy for each segment compared to the SK codes in \cite{song2023nonbinary}.

\begin{theorem}\label{thm:binary}
In a segmented burst-deletion channel, if the input is a binary $(\mathbf{p},\delta)$-sequence of length $n=b\gamma$, which is divided into $\gamma$ non-overlapping segments, each with length $b$, then the original sequence can be decoded successfully and the segment boundaries can be distinguished with $\log b+8\log\log b+o(\log\log b)+r_{q,t}+3$ bits of redundancy per segment.
\end{theorem}

\subsection{Non-binary segmented burst-deletion ECCs}	\label{sub-b}
We now design segmented burst-deletion ECCs for $q$-ary  sequences, where $q>2$. 
In this case, we will use SKQ codes in \cite{song2024new} instead of SK codes in \cite{song2023nonbinary} to encode the input subsequences corresponding to each segment.
Let $\mathbf{x}$ be a $q$-ary $(\mathbf{p},\delta)$-dense sequence  written as $\mathbf{x}=\mathbf{x}_1\mathbf{x}_2\cdots\mathbf{x}_{\gamma}\in\mathcal{P}_q$, where $\mathbf{x}_i$ is the $i$-th segment of $\mathbf{x}$ with $|\mathbf{x}_i|=b,i\in[1,\gamma]$, and $b>2t$. For four non-negative integers $\alpha\in[0,3],\beta\in[0,2b),\eta\in[0,q^N),\zeta\in[0,q^N)$,  where $N=4(\log_q 2\rho)+o(\log_q 2\rho)$ and $\rho=3\delta$,  define
\begin{eqnarray}\label{eqn:Sj}
\mathcal{S}_{\alpha,\beta,\eta,\zeta}^{(j)}\triangleq\left\{\mathbf{s}\in\mathcal{P}_q\subseteq \Sigma_q^n: \mathbf{x}_i \text{ satisfies $f(\mathbf{x}_i)$ with }s_{1},\right.\nonumber\\
\left.	s_{t-1},s_{t},s_{2t-1}\in\Sigma_q\backslash\{j\},s_{b-t+1}=s_{b} \right\},
\end{eqnarray}
where $j\in[0,q)$.
For $j\in[0,q)$, define
\begin{eqnarray*}
(\alpha_j,\beta_j,\eta_j,\zeta_j)=\mathop{\arg\max}\limits_{\substack{0\le \alpha< 4,0\le \beta < 2b,\\ 0\le \eta,\zeta < q^{N}}}|\mathcal{S}^{(j)}_{\alpha,\beta,\eta,\zeta}|.
\end{eqnarray*}

Similar to the binary case, the encoding and decoding of $q$-ary sequences are accomplished by a segment by segment procedure.

\textbf{Encoding:} 
The sequence $\mathbf{x}\in\mathcal{P}_q$ with $n=b\gamma$ is encoded to the concatenation of the codewords in $\mathcal{S}_{\alpha_j,\beta_j,\eta_j,\zeta_j}^{(j)}$, $j\in[0,q)$. Specifically, we encode the first segment to a codeword from $\mathcal{S}^{(0)}_{\alpha_0,\beta_0,\eta_0,\zeta_0}$.
Then,
the $i$-th segment $\mathbf{x}_i,i\in[2,\gamma],$ of length $b$ is encoded into a sequence from $\mathcal{S}^{(j)}_{\alpha_j,\beta_j,\eta_j,\zeta_j}$ if the last bit of the $i-1$-th segment is $j\in[0,q)$. 

\textbf{Decoding:} The decoding procedure is similar to the binary case. 
The received sequence is decoded segment by segment. 
	In the sequel, assume that $\mathbf{x}_{i-1}$ has been recovered successfully, then the last bit of $\mathbf{x}_{i-1}$ is $j,j\in[0,p),$ and the syndrome of $\mathbf{x}_i$ is known, i.e., $f(\mathbf{x}_i)$ with $\alpha_j,\beta_j,\eta_j,\zeta_j$. Denote the  starting position of $\mathbf{x}_i$ by $p_i+1$.

	Denote the received sequence by $\mathbf{y}$ and the $i$-th segment $\mathbf{y}_i=\mathbf{y}(p_i+1:p_i+b)$ as an estimate of $\mathbf{x}_i,i\in[1,\gamma]$. If a burst of $t$-del occurs at the $a$-th component of $\mathbf{x}_i$, the first $t$ symbols of $\mathbf{y}_i$ become $\hat{\mathbf{y}}_i(1:t)=\mathbf{x}_i(1:2t)\backslash\mathbf{x}_i(a:a+t-1)$ for $1\le a\le t$ or $\hat{\mathbf{y}}_i(1:t)=\mathbf{x}_i(1:t)$ for $a>t$.
	Particularly, when $a=1$, $\hat{\mathbf{y}}_i(1:t)=\mathbf{x}_i(t+1:2t)$.	
	Note that it is impossible for $\hat{\mathbf{y}}_i(1:t)$ to be regarded as the last $t$ bits of $\mathbf{x}_{i-1}$ since $\hat{\mathbf{y}}_i(1:t)\ne \mathbf{x}_{i-1}(b-t+1:b)$ following 
	$x_{i,1},x_{i,t+1},x_{i,2t}\ne x_{i-1,b-t+1}=x_{i-1,b}=j$. 
	Therefore, the starting position $p_i+1$ is the first or the $t$-th component of $\mathbf{x}_i$.
	Consequently,
	if  $f(\mathbf{y}_i)\ne f(\mathbf{x}_i)$, according to the syndrome function $f(\mathbf{x}_i)$ and $\mathbf{y}(p_i+1:p_i+b-t)$,  the $i$-th segment can be recovered
	and the starting position of  $\mathbf{x}_{i+1}$ is found as $p_i+b-t+1$. 
	Otherwise, we deduce that $\mathbf{x}_i=\mathbf{y}_i$ and determine the starting position of $\mathbf{x}_{i+1}$, as described in Lemma \ref{decoding}.

	\begin{lemma}\label{decoding}
		Let a non-binary $(\mathbf{p},\delta)$-dense sequence $\mathbf{x}=\mathbf{x}_1\mathbf{x}_2\cdots\mathbf{x}_{\gamma}$ be the input and $\mathbf{y}$ be the output of a segmented burst-deletion channel.
		Assume that the $i-1$-th segment of $\mathbf{x}$ is recovered successfully and the starting position  of $\mathbf{x}_i$ is $p_i+1$.	
		Denote $\mathbf{y}_i=\mathbf{y}(p_i+1:p_i+b)$. If $f(\mathbf{y}_i)=f(\mathbf{x}_i)$, then $\mathbf{x}_i=\mathbf{y}_i$ and the starting position of the $i+1$-th segment is $p_i+b+1$.
	\end{lemma}
	\begin{proof}
		Without loss of generality, suppose that the last bit of $\mathbf{x}_{i-1}$ is $0$, then the syndrome of $\mathbf{x}_i$ is $f(\mathbf{x}_i)$ with $(\alpha_0,\beta_0,\eta_0,\zeta_0)$. The decoding for the case with the last bit being $j,j\in[1,q)$,  syndrome $f(\mathbf{x}_i)$ with $(\alpha_j,\beta_j,\eta_j,\zeta_j)$ is similar.
		
		Following the encoding approach and \eqref{eqn:Sj}, if there is no burst of $t$-del in $\mathbf{x}_i$, $x_{p_i+1}$ should bot be $0$.
		Therefore, if  $y_{p_i+1}=0$, it means that the initial $t$ bits of $\mathbf{x}_i$  are deleted.
		If $f(\mathbf{y}_i)=f(\mathbf{x}_i)$,  there exist two different sequences $\mathbf{x}_i$ and $\mathbf{y}_i=\mathbf{y}(p_i+1:p_i+b)=\mathbf{x}_i(t+1:b)\cup\hat{\mathbf{y}}_{i+1}(1:t)$  with
		the same syndrome $(\alpha_0,\beta_0,\eta_0,\zeta_0)$ and a common subsequence of length $b-t$  obtained by deleting $t$ consecutive symbols. This implies
		$\mathcal{B}_t^D(\mathbf{y}_i)\cap\mathcal{B}_t^D(\mathbf{x}_i)\ne \varnothing$. Hence, there is a contradiction.

		
		In the following, we consider the case with $y_{p_i+1}\ne 0$ by analyzing the following five cases shown in Figure \ref{fig}.
		\begin{itemize}
			\item [1.]	There is no burst of $t$-del in the $i-1$-th segment, and one burst of $t$-del in the $i$-th segment, which does not destroy the first bit of $\mathbf{x}_i$ (Case 1). Since we have recovered $\mathbf{x}_{i-1}$ and $\hat{\mathbf{y}}_i(1:t)$ can not be regarded as the last $t$ bits of $\mathbf{x}_{i-1}$, thus $y_{p_i+1}=x_{i,1}$. 
			Following this, if $f(\mathbf{y}_i)=f(\mathbf{x}_i)$,  there exist two different sequences $\mathbf{x}_i$ and $\mathbf{y}_i=\mathbf{y}(p_i+1:p_i+b)=\mathbf{x}_i(1:a-1)\cup\mathbf{x}_i(a+t:b)\cup\hat{\mathbf{y}}_{i+1}(1:t)$  with
			the same syndrome $(\alpha_0,\beta_0,\eta_0,\zeta_0)$ and a common subsequence of length $b-t$  obtained by deleting $t$ consecutive symbols, where $a\in[2,b-t]$ is the  position of burst-deletion in $\mathbf{x}_i$. Note that the fact of $\hat{\mathbf{y}}_{i+1}(1:t)\ne\mathbf{x}_i(b-t+1:b)$ ensures that  $\mathbf{x}_i$ and $\mathbf{y}_i$ are different.
			This implies
			$\mathcal{B}_t^D(\mathbf{y}_i)\cap\mathcal{B}_t^D(\mathbf{x}_i)\ne \varnothing$, contradicting the fact that $f(\mathbf{x})$ with $(\alpha_0,\beta_0,\eta_0,\zeta_0)$ can correct one burst of $t$-del.

			\item [2.]
			There is one burst of $t$-del in the $i-1$-th segment and one in the $i$-th segment, respectively, which does not destroy the first bit of the latter (Cases 2-3).
			The burst-deletion of the $i-1$-th segment can be corrected successfully. After that, the proof is reduced to that of Case 1.
			
			\item [3.]
			There is one burst of $t$-del in the $i-1$-th segment and one affecting the initial $t$ bits of the $t$-th segment (Cases 4-5).
			The burst-deletion of the $i-1$-th segment can be corrected successfully.
			Hence, $\mathbf{y}_i=\mathbf{y}(p_i+1:p_i+b)=\mathbf{x}_i(t+1:b)\cup\hat{\mathbf{y}}_{i+1}(1:t)$.
			If $f(\mathbf{y}_i)=f(\mathbf{x}_i)$,  there exist two different sequences $\mathbf{x}_i$ and $\mathbf{y}_i$  with
			the same syndrome $(\alpha_0,\beta_0,\eta_0,\zeta_0)$ and a common subsequence of length $b-t$  obtained by deleting $t$ consecutive symbols. This implies
			$\mathcal{B}_t^D(\mathbf{y}_i)\cap\mathcal{B}_t^D(\mathbf{x}_i)\ne \varnothing$. Hence, there is a contradiction.
		\end{itemize}
		
		Following the processes above, if $f(\mathbf{y}_i)=f(\mathbf{x}_i)$, then we know $\mathbf{y}_i=\mathbf{x}_i$, resulting in the starting position of $\mathbf{x}_{i+1}$ being $p_i+b+1$, completing the proof.	
	\end{proof}

	In the sequel, we analyze the cardinality and rate of the designed codes.
From \eqref{eqn:Sj},
there are $q^{b-5}(q-1)^4$ sequences of length $b$ satisfying the restrictions.
Each of these sequences belongs to one of the sets $\mathcal{S}_{\alpha_j,\beta_j,\eta_j,\zeta_j}^{(j)},j\in[0,q)$. Thus, the size of the largest one among these  sets is lower bounded as
\begin{eqnarray}\label{eqn:Mq}
	M\ge {q^{b-5}(q-1)^4 \over 4\cdot 2b\cdot q^{2N}},
\end{eqnarray}

Taking logarithms of both sides, and applying \eqref{eqn:rate}, the rate of the designed codes is lower bounded by
\begin{eqnarray*}
	R\ge \log q-{|f(\mathbf{x})|\over b}-{5\log q\over b}+{4\log (q-1)\over b},
\end{eqnarray*}
where $f(\mathbf{x})$ is defined in Lemma \ref{lemma:burst} for non-binary sequences.

We see that the rate penalty with respect to  SKQ codes is at most $(5\log q-4\log (q-1))/ b$ bits per symbol.
Thus, the redundancy of the new codes is $\log b+8\log\log b+o(\log\log b)+r_{q,t}+5\log q-4\log (q-1)$ bits.
In other words, to correct one burst of $t$-del per segment over the $q$-ary segmented burst-deletion channel, we only need to add $5\log q-4\log (q-1)$ bits more redundancy for each segment compared to the SKQ codes in \cite{song2024new}.
\begin{theorem}\label{thm:q-ary}
	In a segmented burst-deletion channel, if the input is a non-binary $(\mathbf{p},\delta)$-sequence of length $n=b\gamma$, which is divided into $\gamma$ non-overlapping segments,  each with length $b$, then the original sequence can be decoded successfully and the segment boundaries can be distinguished with $\log b+8\log\log b+o(\log\log b)+r_{q,t}+5\log q-4\log (q-1)$ bits of redundancy per segment.
\end{theorem}

		\begin{remark}
	The function $f(\mathbf{x})$ applied in the new constructions could be any one burst of $t$-del ECCs, not limited to SK codes and SKQ codes, and the encoding/decoding algorithms can be developed analogously.
	\end{remark}

Since our codes are designed for the $(\mathbf{p},\delta)$-dense sequences, we need to analyze the redundancy involved in encoding a sequence into a $(\mathbf{p},\delta)$-dense sequence.
	\begin{lemma}\label{lem:dense}
		For given integers $q,b,\gamma,t,n=b\gamma,\delta=2tq^{2t}\lceil \log n\rceil$ and $\mathbf{p}=0^t1^t$, if $\mathbf{x}=\mathbf{x}_1\mathbf{x}_2\cdots\mathbf{x}_{\gamma}\in\Sigma_q^n$ is a $(\mathbf{p},\delta)$-dense sequence, then each segment $\mathbf{x}_i$ is also a $(\mathbf{p},\delta)$-dense sequence, where $i\in[1,\gamma]$.
	\end{lemma}
	\begin{proof}
		According to the definition of $(\mathbf{p},\delta)$-dense sequence, we know there is at least one $\mathbf{p}$ in the  substrings of length $\delta$ in $\mathbf{x}$.
		That means, if $b\ge\delta$, there is at least one $\mathbf{p}$ in the substrings of  length $\delta$ in $\mathbf{x}_i,i\in[1,\gamma]$, which states that each $\mathbf{x}_i$ is a $(\mathbf{p},\delta)$-dense sequence.
		If $b<\delta$, it is clear that each $\mathbf{x}_i$ is also a $(\mathbf{p},\delta)$-dense sequence, as the condition for density is trivially satisfied.
	\end{proof}

	Following Remark \ref{rem:burst}, another $3$ bits and one bit of redundancy is necessary for binary and non-binary $(\mathbf{p},\delta)$-dense sequence $\mathbf{x}$, respectively.
Thus, the average redundancy per segment is $\log b+8\log\log b+o(\log\log b)+r_{q,t}+3+3/\gamma$ bits for binary segmented ECCs proposed in Subsection \ref{sub-a}, and $\log b+8\log\log b+o(\log\log b)+r_{q,t}+5\log q-4\log (q-1)+1/\gamma$ bits for the non-binary case in \ref{sub-b}.

	\subsection{Comparisons}
In \cite{sima2020syndrome}, Sima \textit{et al.} constructed the ECCs, which can be used to correct at most $\gamma$ bursts of $t$-del for binary sequences of length $n$ with $4\gamma\log n+o(\log n)$ bits of redundancy. This result can easily be extended to $q$-ary sequences.
Particularly, in segmented burst-deletion channels,
the $\gamma$ bursts are located in different segments.
For any $q$-ary sequence $\mathbf{x}\in\Sigma_q^n$, define 
\begin{eqnarray*}\label{eqn:N_T,t}
	\mathcal{N}_{q,\gamma,t}(\mathbf{x})\triangleq\{\mathbf{x}'\in \Sigma_q^n:~\mathbf{x}'\ne \mathbf{x}, ~\mathcal{B}_{\gamma,t}^{D}(\mathbf{x}')\cap \mathcal{B}_{\gamma,t}^{D}(\mathbf{x})
	\ne \varnothing\}.
\end{eqnarray*}

	%

Each codeword in $\mathcal{N}_{q,\gamma,t}(\mathbf{x})$ can be obtained following the two steps below:
\begin{itemize}
	\item Delete at most $\gamma$ substrings of length $t$ from $\mathbf{x}$ with no more than one substring in each segment, resulting in $\mathbf{y}\in\Sigma_q^{m},m\in[n-\gamma t,n]$. There are at most $b-t+2$ possibilities per segment, leading to at most $(b-t+2)^{\gamma}$ possibilities for $\mathbf{y}$.
	\item For each $\mathbf{y}$, insert at most $\gamma$ sequences of length $t$ into $\mathbf{y}$, with no more than one in each segment, to obtain a sequence $\mathbf{x}\in\Sigma_q^{n}$.  There are at most $((b-t+2)q^{t})^{\gamma}$ possibilities for $\mathbf{x}$. 
\end{itemize}

These two steps above result in
\begin{eqnarray*}
	|\mathcal{N}_{q,\gamma,t}(\mathbf{x})|
	\le (b-t+2)^{\gamma}\cdot ((b-t+2)q^{t})^{\gamma}
	\le b^{2\gamma}q^{\gamma t}.
\end{eqnarray*}

Following the syndrome compression technique in \cite{sima2020syndrome}, one can construct a function with 
$2\log_q |\mathcal{N}_{q,\gamma,t}(\mathbf{x})|+o(\log_q b)=4\gamma \log_q b+o(\log_q b)$ symbols of redundancy to correct at most $\gamma$ bursts of $t$-del  for segmented burst-deletion channels. 
In contrast to the codes in \cite{sima2020syndrome}, the redundancy of the codes designed in Sections \ref{sub-a} and \ref{sub-b} per segment is  $\log b+8\log\log b+o(\log\log b)+r_{q,t}+3$ bits for binary  and $\log b+8\log\log b+o(\log\log b)+r_{q,t}+5\log q-4\log (q-1)$ bits for non-binary cases, respectively. For both cases, the overall redundancy for encoding the entire sequence  $\mathbf{x}$ is $\gamma\log_q b+o(\log_q b)$ symbols if we take logarithms with base $q$ on the rate, which improves the redundancy of ECCs in \cite{sima2020syndrome} by a factor of $4$. It is noteworthy that the approach in \cite{sima2020syndrome} is capable of correcting general $t$-error patterns, comprising any combination of $t_1$-del, $t_2$-ins, and $t_3$-sub, as long as $t=t_1+t_2+t_3$. By contrast, the codes developed in this paper are only limited to at most one burst of $t$-del per segment for the segmented burst-deletion channel.

In another related work \cite{yi2024error}, the authors provide two segmented burst indel codes, namely
binary marker MDS (BM-MDS) codes and binary marker de Bruijn symbol MDS (BM-DB-MDS) codes, which can correct up to $t_1$-del or up to $t_2$-ins in each segment of length $b$. Therein, $b=u(t_1+t_2+2)$, where $1/u~(u>2)$  represents the rough proportion of the maximum length of burst insertions or deletions allowed in a codeword. 
In these constructions, the original $k$ information symbols are first encoded by a $(b,k)$ MDS code over $\mathbb{F}_q,q\ge 2^{\lfloor1+ \log b\rfloor}$.
Subsequently, a binary marker sequence is applied for detection of the length and type (insertion/deletion) of burst errors. Moreover, the de Bruijn sequence is utilized to precisely localize error positions in BM-DB-MDS codes. 
In this way, the redundancies of  BM-MDS codes and BM-DB-MDS codes are $b\log b/u+\log b$ and $((\lambda+2u)b\log b)/(2u(\lambda+1))+\log b$ bits, respectively, which scale on the order of $O(b\log b)$, where  $\lambda=(\log q)/\lfloor 1+ \log b\rfloor$ is the shortening factor of the applied MDS codes.

In contrast to \cite{yi2024error}, the segmented burst-deletion codes proposed in Section \ref{sec:deletion} operate over any field size $q\ge 2$, eliminating the requirement of $q\ge 2^{\lfloor 1+ \log b\rfloor}$. The new codes achieve a redundancy of order $O(\log b)$, which significantly improve the results in \cite{yi2024error}. However, the codes proposed in this paper  can only correct at most one burst of $t$-del in each segment, while the one in \cite{yi2024error} can correct one burst of up to $t_1$-del/up to $t_2$-ins per segment.


\section{Conclusions}\label{sec:conclusions}
We constructed $q$-ary ($q\ge 2$) ECCs for segmented burst-deletion channels, in which at
most one burst of $t$-del occurs within each segment,
achieving a redundancy per segment that scales as $O(\log b)$.
The codes were constructed by applying existing one-burst deletion codes, with additional constraints that enable the determination of segment boundaries during the decoding process from the received sequence.
	Extensions of the designed codes to segmented-insertion and segmented-edit channels, and development of segmented burst indel/edit ECCs for correcting at most one burst of up to $t_1$-del or $t_2$-ins per segment with lower redundancy represent interesting future directions.
	\ifCLASSOPTIONcaptionsoff
	\newpage
	\fi

	\bibliographystyle{IEEEtran}
	\bibliography{myreference}

\end{document}